\documentclass[aps,twocolumn,superscriptaddress,floatfix,longbibliography]{revtex4-2}
\usepackage{amsmath}
\usepackage{amssymb}
\usepackage{graphicx}
\usepackage{dcolumn}
\usepackage{bm}
\usepackage{verbatim}
\usepackage{comment}
\usepackage{stmaryrd}
\usepackage{color}
\usepackage{grffile}
\usepackage{lipsum}
\usepackage{enumitem}
\usepackage[colorlinks,linkcolor=blue,citecolor=blue,urlcolor=blue]{hyperref}

\begin{document}

\title{Localization and mobility edges in non-Hermitian disorder-free lattices}

\author{Rui Qi}
\affiliation{Beijing National Laboratory for Condensed Matter Physics, Institute of Physics, Chinese Academy of Sciences, Beijing 100190, China}
\affiliation{School of Physical Sciences, University of Chinese Academy of Sciences, Beijing, 100049, China}
\author{Junpeng Cao}

\affiliation{Beijing National Laboratory for Condensed Matter Physics, Institute of Physics, Chinese Academy of Sciences, Beijing 100190, China}
\affiliation{School of Physical Sciences, University of Chinese Academy of Sciences, Beijing, 100049, China}
\affiliation{Songshan lake Materials Laboratory, Dongguan, Guangdong 523808, China}
\affiliation{Peng Huanwu Center for Fundamental Theory, Xian 710127, China}
\author{Xiang-Ping Jiang}
\email{2015iopjxp@gmail.com}
\affiliation{Zhejiang Lab, Hangzhou 311121, China}

\date{\today}

\begin{abstract}
The non-Hermitian skin effect (NHSE) is a significant phenomenon observed in non-Hermitian systems under open boundary conditions, where the extensive bulk eigenstates tend to accumulate at the lattice edges. In this article, we investigate how an electric field affects the localization properties in a non-Hermitian mosaic Stark lattice, exploring the interplay between the Stark localization, mobility edge (ME), and the NHSE induced by nonreciprocity. We analytically obtain the Lyapunov exponent, the phase transition points and numerically calculate  the density distributions and the spectral winding number. We reveal that in the nonreciprocal Stark lattice with the mosaic periodic parameter $\kappa=1$, there exists a critical electric field strength that describes the transition of the existence-nonexistence of NHSE and is inversely proportional to the lattice size. This transition is consistent with the real-complex transition and topological transition characterized by spectral winding number under periodic boundary conditions. In the strong fields, the Wannier-Stark ladder is recovered, and the Stark localization is sufficient to suppress the NHSE. When the mosaic period $\kappa=2$, we show that the system manifests an exact non-Hermitian ME and the skin states are still existing in the strong fields, in contrast to the gigantic field can restrain the NHSE in the $\kappa=1$ case. Moreover, we further study the expansion dynamics of an initially localized state and dynamically probe the existence of the NHSE and the non-Hermitian ME. These results could help us to control the NHSE and the non-Hermitian ME by using electric fields in the disorder-free systems.
\end{abstract}

\maketitle

\section{Introduction}

Anderson localization \cite{anderson1958absence,abrahams1979scaling,evers2008anderson} is a universal and significant quantum phenomenon, which has attracted vast amounts of research in the condensed-matter physics. In one and two dimensions, infinitesimal quenched disorder can localize the all single-particle eigenstates, and the localization transitions do not happen. For one dimensional (1D) systems, when the random disorder is replaced with quasiperiodic disorder, the system can manifest the delocalization-localization transition. 
The most famous quasiperiodic model is Aubry-Andr{\'e}-Harper (AAH) model \cite{harper1955single,aubry1980analyticity}, and its analytical localization transition point is determined by the self-duality symmetry. In addition, some variants of AAH model breaking the self-duality exhibit the single-particle mobility edge (ME), which is a critical energy separating the extended and localized eigenstates \cite{li2017mobility,rossignolo2019localization,biddle2010pre,wang2020one,duthie2021self,zhang2022lyapunov,wang2021duality,gonccalves2022hidden,liu2022general,zhou2022exact,wang2023two,vu2023generic,liu2022general}. 

However, in the absence of the random or quasiperiodic disorder, an electric field applied to a 1D disorder-free lattice also induce the localization transition, usually dubbed as the Wannier-Stark localization \cite{wannier1962dynamics,fukuyama1973tightly,emin1987existence}. The Wannier-Stark localization has some different properties from the localization induced by random or quasiperiodic disorder, such as the equally-spaced energy spectrum (Wannier-Stark ladder) in the localized phase and Bloch oscillation \cite{hartmann2004dynamics}. Moreover, the single-particle ME without disorder has been found in the mosaic Stark model where on-site potentials are inlaid in the lattice with equally spaced sites \cite{dwiputra2022single}.

Recently, the interplay of the non-Hermiticity and Anderson localization has attracted much attention and been extensively studied in both quenched disorder and quasiperiodic systems \cite{longhi2019topological,longhi2019metal,jiang2019interplay,zeng2020topological,zeng2020winding,tzortzakakis2020non,liu2020non,liu2020generalized,zeng2020topological,schiffer2021anderson,xu2020topological,zhang2020skin,liuT2020non,tang2021localization,schiffer2021anderson,liu2021localization,liu2021exactb,liu2021exacta,zhai2020many,hamazaki2019non,tang2021localization,cai2022localization,sarkar2022interplay,jiang2021mobility,jiang2021non,jiang2023mobility}. In the non-Hermitian systems, there exist some exotic phenomena without Hermitian counterpart. For example, a significant phenomenon induced by the nonreciprocal hopping is the non-Hermitian skin effect (NHSE) \cite{yao2018edge,kunst2018biorthogonal,longhi2019probing,song2019non,yokomizo2019non,yi2020non,yang2020non,liu2020helical,gou2020tunable,okuma2020topological,zhang2020correspondence,longhi2021phase,lu2021magnetic,claes2021skin,li2021quantized,li2020critical,guo2021exact,guo2022theoretical,zeng2022real,longhi2022self,longhi2022non,wang2022non,suthar2022non,alsallom2022fate,kawabata2022many,zhang2022symmetry,kawabata2023entanglement}, i.e., the 
extensive bulk eigenstates are exponentially localized at the boundaries of the systems under open boundary conditions (OBC). Meanwhile, the single-particle MEs have been generalized to the non-Hermitian quasiperiodic lattices in recent years. However, the interplay between the NHSE and the Stark localization as well as the ME in disorder-free systems is still rarely studied \cite{peng2022manipulating,wang2022tightly,zhang2022engineering}. 

In this work, we propose a non-Hermitian mosaic Stark model to investigate the effect of electric fields on NHSE and non-Hermitian ME. We reveal that in the model with the mosaic perodic parameter $\kappa = 1$, there exists a critical electric field strength that describes the analytical transition of the existence-nonexistence of NHSE and is inversely proportional to the lattice size. This transition is consistent with the real-complex transition and topological transition characterized by spectral winding number under periodic boundary conditions (PBC). For the mosaic period $\kappa = 2$, we find that a non-Hermitian ME emerges and the Stark localized states appear gradually above a critical weak field. Especially, the skin states are still existing in the strong fields under OBC, in contrast to the $\kappa = 1$ case the large fields can suppress the NHSE. Hence, the manipulation methods of the electric fields on the NHSE and the non-Hermitian ME are abundant, and can be easily applied to the mosaic disorder-free systems.

The rest of the work is organized as follows.
We first introduce the non-Hermitian mosaic Stark model in Sec. \ref{section2}. In Sec. \ref{section3}, we analytically calculate the Lyapunov exponent (LE) of the mosaic Stark model using the Avila's global theory. We show that the transition of the existence-nonexistence of the NHSE is always accompanied with the real-complex and topological transition under PBC. In Sec. \ref{section4}, we obtain the exact ME without disorder in the non-Hermitian mosaic Stark model with $\kappa=2$.  In Sec. \ref{section5}, we also study the expansion dynamics to detect the NHSE and demonstrate the existence of the non-Hermitian ME. A brief conclusion is given in Sec. \ref{sec:conclusion}.

\begin{figure*}
\includegraphics[width=0.8\textwidth]{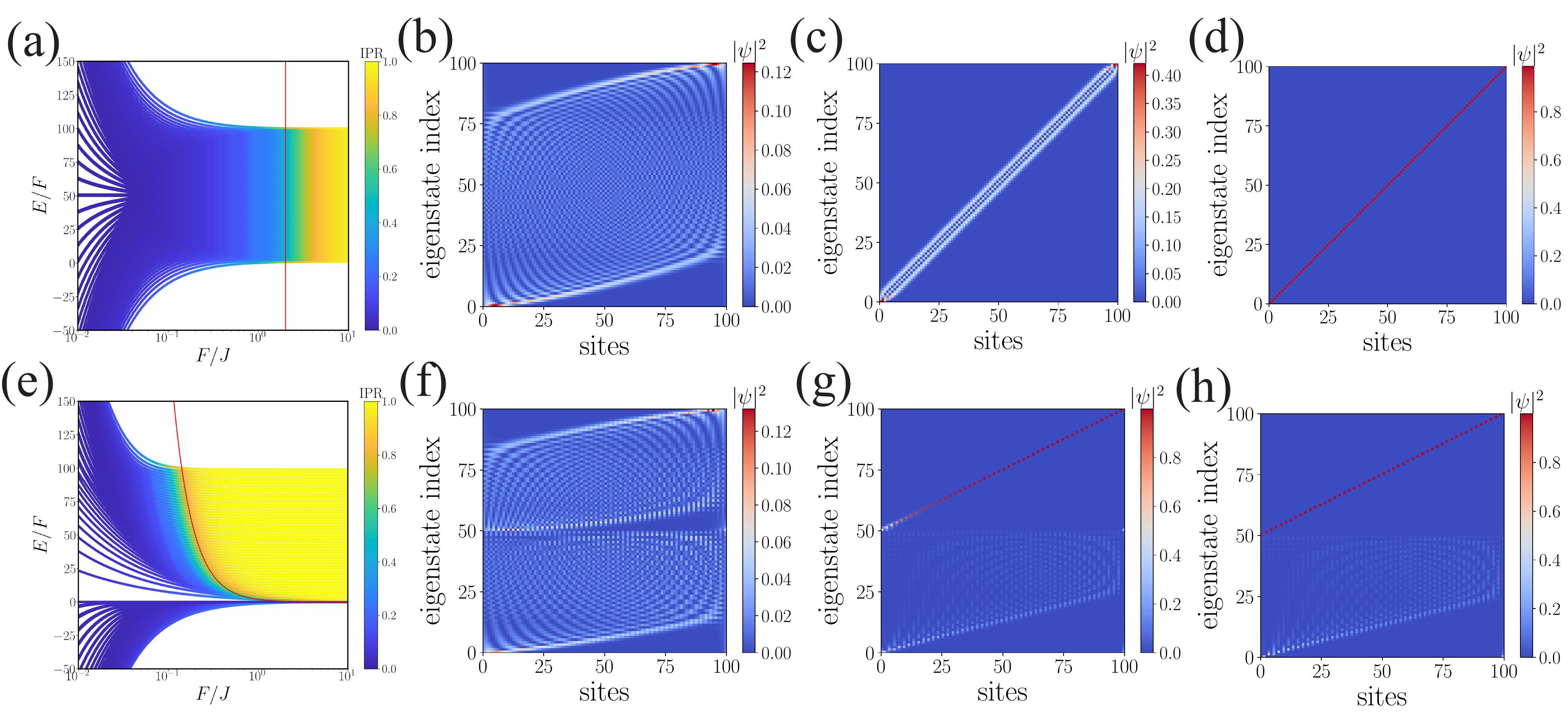}
\caption{The $\rm{IPR}$ versus $F/J$ for Hermitian Stark model with $L=100$ under PBC in (a) and (e) corresponding to the case of $\kappa = 1$ and $\kappa = 2$. Here, the yellow and blue colors denote the localized and extended states, respectively. The red solid lines in (a) and (e) represent the transition point $F/J=2$ and ME, respectively. The density distribution of all eigenstates for Hermitian Stark model under PBC with system size $L=100$ from left to right corresponding to $F/J = 0.01$, $0.5$, and $10$ in (b)-(c) and (f)-(h). The top row (a)-(d) and the bottom row (e)-(h) correspond to the case of $\kappa = 1$ and $\kappa =2$, respectively.}
\label{fig1}
\end{figure*}
\section{The model Hamiltonian with nonreciprocal hopping}\label{section2}
We consider a 1D non-Hermitian mosaic Stark model, which is described by the Hamiltonian
\begin{equation}\label{model}
	H=\sum_{n}J(e^{g}c_n^{\dagger}c_{n+1} + e^{-g}c_{n+1}^{\dagger}c_n)+ \sum_{n} V_{n} c_n^{\dagger}c_n,
\end{equation}
where $c^{\dagger}_n$ and $c_n$ are the fermionic creation and annihilation operators at $n$-th site, respectively. $J e^{\pm g}$ are the nearest-neighbor nonreciprocal hopping amplitudes and $g$ is non-Hermitian parameter. Here, $V_n$ is the onsite potential at site $n$ with the form
\begin{eqnarray}\label{eq:Disord} 
V_n=\left\{ 
\begin{array}{cl}
	F n\cos{\theta}, & n=\kappa j  ,\\
	0,& \mathrm{otherwise},\\
\end{array} 
\right.
\end{eqnarray}
where $F$ and $\kappa$ denote the electronic field and the mosaic periodic parameter, respectively. Because the onsite potential occurs with interval $\kappa$ in this model, we introduce a supercell with the nearest-$\kappa$ sites. If the supercell number of the system is denoted as $N$, i.e. $j = 1,2,...,N$, the lattice length will be $L=\kappa N$. 

If $\kappa=1$ and $g = 0$, this model is reduced to the Wannier-Stark model where the eigenstates are described by  $|\Psi_i\rangle= \sum_n \mathcal{J}_{n-i}(2J/F)|n\rangle$ \cite{wannier1962dynamics,fukuyama1973tightly}, where $|n\rangle$ is the computational basis and  $\mathcal{J}_{n-i}(2J/F)$ is the $(n-i)$-th Bessel function of the first kind of $J$ and $F$. When $F/J< 2$, all the eigenstates are extended. While $F/J>2$, all the eigenstates are localized and form the equally-spaced energy spectrum, i.e., Wannier-Stark ladders. Therefore, there is no mobility edge in the Wannier-Stark model, as shown in Fig. \ref{fig1}(a). 

Here, we show the inverse participation ratio ($\rm{IPR} $) of different eigenstates with the eigenenergies $E$ versus $F/J$ in Figs. \ref{fig1}(a) and (e), which can be used to distinguish extended and localized states. The $\rm{IPR}$ of $i$-th eigenstate $|\Psi_i\rangle$ is defined as 
\begin{eqnarray}
\mathrm{IPR}^i = \sum_{m=1}^{L} |\psi_m^i|^{4},
\end{eqnarray}
where $\psi_m^i$ represents $m$-th element of $i$-th normalized eigenstate. For a localized state, ${\rm {IPR}}^i \sim 1$, while for a extended state, ${\rm {IPR}}^i \sim 1/L$. 

We show the $\rm IPR$ and the energy spectrum versus $F/J$ for the $\kappa = 1$ case under PBC in Fig. \ref{fig1}(a). We find that all eigenstates are extended states for $F/J<2$, while the eigenstates are localized for $F/J>2$. Note that we set lattice size $L=100$ in the numerical calculations. In Figs. \ref{fig1}(b)-(d), we plot the density distribution of all eigenstates of Hermitian Stark model for different constant forces $F/J = 0.01,0.1$, and $10$. When $F/J<2$, the extended eigenstates are distributed over a finite region, which are different from the random disorder and quasiperiodic systems where the particles are distributed uniformly in the extended phase. Whereas $F/J>2$, the state is localized where the single particle only locates at one site.

For $\kappa =2$ case in Fig. \ref{fig1}(e), we show that there exist fully extended phases when $F/J \lesssim 0.1$. When $F/J \simeq 0.1$, extended states begin to appear accompanied by a mobility edge, which is denoted by the red solid line in Fig. \ref{fig1}(e). After introducing a mosaic structure, i.e., $\kappa>1$, extended states appear earlier compared with the case of $\kappa =1$. Moreover, the system no longer has fully localized phases but mobility edge phases with the increasing of $F/J$. In Figs. \ref{fig1}(f)-(h), we plot the density distribution of all eigenstates for the Hermitian Stark model (\ref{model}) with $g = 0$ and $\kappa =2$. We clearly see when $F/J = 10 $, half of the eigenstates above the mobility edge are extended and the others are localized in Fig. \ref{fig1}(h), which is different with the case of $\kappa=1$.

\section{Lyapunov exponent, non-Hermitian skin effect and real-complex transition}
\label{section3}
In the following, we consider the non-Hermitian mosaic Stark lattice (\ref{model}) and  analytically calculate the LE by applying a similar transformation and Avila's global theory \cite{avila2015global}. We transform the non-Hermitian Hamiltonian (\ref{model}) into a Hermitian Hamiltonian $H^{\prime}$ via a similarity transformation under OBC,
\begin{equation}\label{model_NH}
	H^{\prime}=SH(g)S^{-1},
\end{equation}
where the similar matrix $S= {\rm diag} (e^{-g}, e^{-2g}, \cdots, e^{-Lg})$ is diagonal, and $H^{\prime}=H(g=0)$ is the Hermitian Hamiltonian. The LE of $H^{\prime}$ can be obtained via Avila's global theory. Note that the spectrum of $H$ and $H^{\prime}$ are same under the similar transformation. Here, we set $J = 1$ for simplicity and take $|\Psi^{\prime} \rangle=\sum_n \psi_{n}^{\prime} |n \rangle$ into the eigenequation $H^{\prime} | \Psi^{\prime} \rangle = E| \Psi^{\prime} \rangle$. So we get the Shr${\rm \ddot o}$dinger equation in the form of $E\psi_{n}^{\prime} = \psi_{n+1}^{\prime} + \psi_{n-1}^{\prime} + V_{n}\psi_{n}^{\prime}$, then the transfer matrix equation is written as
\begin{equation}
\left(\begin{matrix}
\psi_{n+1}^{\prime} \\ \psi_n^{\prime}
\end{matrix}\right) = T_n
\left(\begin{matrix}
\psi_{n}^{\prime} \\ \psi_{n-1}^{\prime}
\end{matrix}\right),
\end{equation}
where the transfer matrix $T_n$ is given by
\begin{align}\label{transfer}
T_{n} &=
\left(\begin{matrix}
E-V_n & -1 \\
1 & 0 
\end{matrix}\right).  
\end{align}
The transfer matrix in a quasicell can be written as
\begin{align}
T_{\kappa,j}(E,\theta)&=\left(\begin{matrix}
E-F\kappa j \cos{\theta} & -1 \\
1 & 0 
\end{matrix}\right)
\left(\begin{matrix}
E & -1 \\
1 & 0 
\end{matrix}\right)^{\kappa-1}.
\end{align}
Using matrix eigendecomposition, we can calculate the ($\kappa-1$)-th power of the matrix as
\begin{align}\label{transfer1}
\left(\begin{matrix}
E & -1 \\
1 & 0 
\end{matrix}\right)^{\kappa-1}
=\left(\begin{matrix}
a_{\kappa} & -a_{\kappa-1} \\
a_{\kappa-1} & -a_{\kappa-2} 
\end{matrix}\right)
\end{align}
with the coefficients

\begin{equation}\label{a}
\small
a_\kappa = \frac{1}{\sqrt{E^2-4}}\left[\left(\frac{E+\sqrt{E^2-4}}{2}\right)^\kappa - \left(\frac{E-\sqrt{E^2-4}}{2}\right)^\kappa\right].
\end{equation}

Now we analytically calculate the LE based on the transfer matrix, which can be represented as 
\begin{equation}\label{LE}
\gamma \left( E\right) =\lim_{L\rightarrow \infty }\frac{1}{2\pi L}\int \ln\left\vert \left\vert T_{L}\left( E,\theta \right) \right\vert \right\vert d\theta ,
\end{equation}
where $\left\vert \left\vert \cdot \right\vert \right\vert $ represents the norm of
a matrix and
\begin{equation}\label{fullLE}
T_{L}\left( E,\theta \right) =\prod_{j=1}^{N}T_{\kappa, j}\left(E, \theta\right) .
\end{equation}
is the transfer matrix for all sites. Based on the Avila’s theory and the discussions in Refs. \cite{wang2020one,dwiputra2022single,liu2021exacta}, we know that if $T_{k,j}\left( E,\theta \right)$ has a holomorphic extension to the neighborhood of $\rm Im(\theta)$, we can define $T_{\kappa,j} (E,\theta)=T_{\kappa,j} (E,\theta+i\epsilon)$. Then letting $\epsilon \rightarrow \infty $, the transfer matrix becomes
\begin{equation}
T_{\kappa,j}\left(E, \theta+i\epsilon \right) =\frac{j\kappa}{2} e^{-i\theta}e^{|\epsilon|}\left(
\begin{array}{cc}
-Fa_{\kappa}& Fa_{\kappa-1} \\
0 & 0
\end{array}
\right) + O(1).
\end{equation}
Thus, based on Eq. (\ref{fullLE}) and a direct computation, we get $\|T_L(E)\|=N!|\frac{F}{2}e^{\epsilon}\kappa a_{\kappa}|^N$. Within the Stirling's approximation, $\ln N! \approx N \ln N - N$ and $\epsilon \rightarrow \infty$, we have 
\begin{equation}\label{eq13}
\kappa\gamma_{\epsilon}(E)=\ln\left|\frac{F}{2}a_{\kappa}\right|+ \ln(\kappa N) + |\epsilon| -1.
\end{equation}
Avila's global theory also shows that as a function of $\epsilon$, the LE is a convex piecewise linear function with integer slopes. As we can see in the above, the slope of $\gamma_{\epsilon}(E)$ with respect to $\epsilon$ for $\epsilon \rightarrow \infty$ is always $1$. However, the slope might be $1$ (when $E$ is in the spectrum) or $0$ (when $E$ is not) in the neighborhood of $\epsilon \rightarrow 0^{+}$. Thus, if $E$ lies in the spectrum of the Hamiltonian $H(g=0)$, we obtain $\kappa\gamma_{\epsilon}(E)={\rm{max}}\{\ln\left|\frac{F}{2}a_{\kappa}\right| + \ln(\kappa N)-1,0\}$. 
\begin{figure*}
\includegraphics[width=0.8\textwidth]{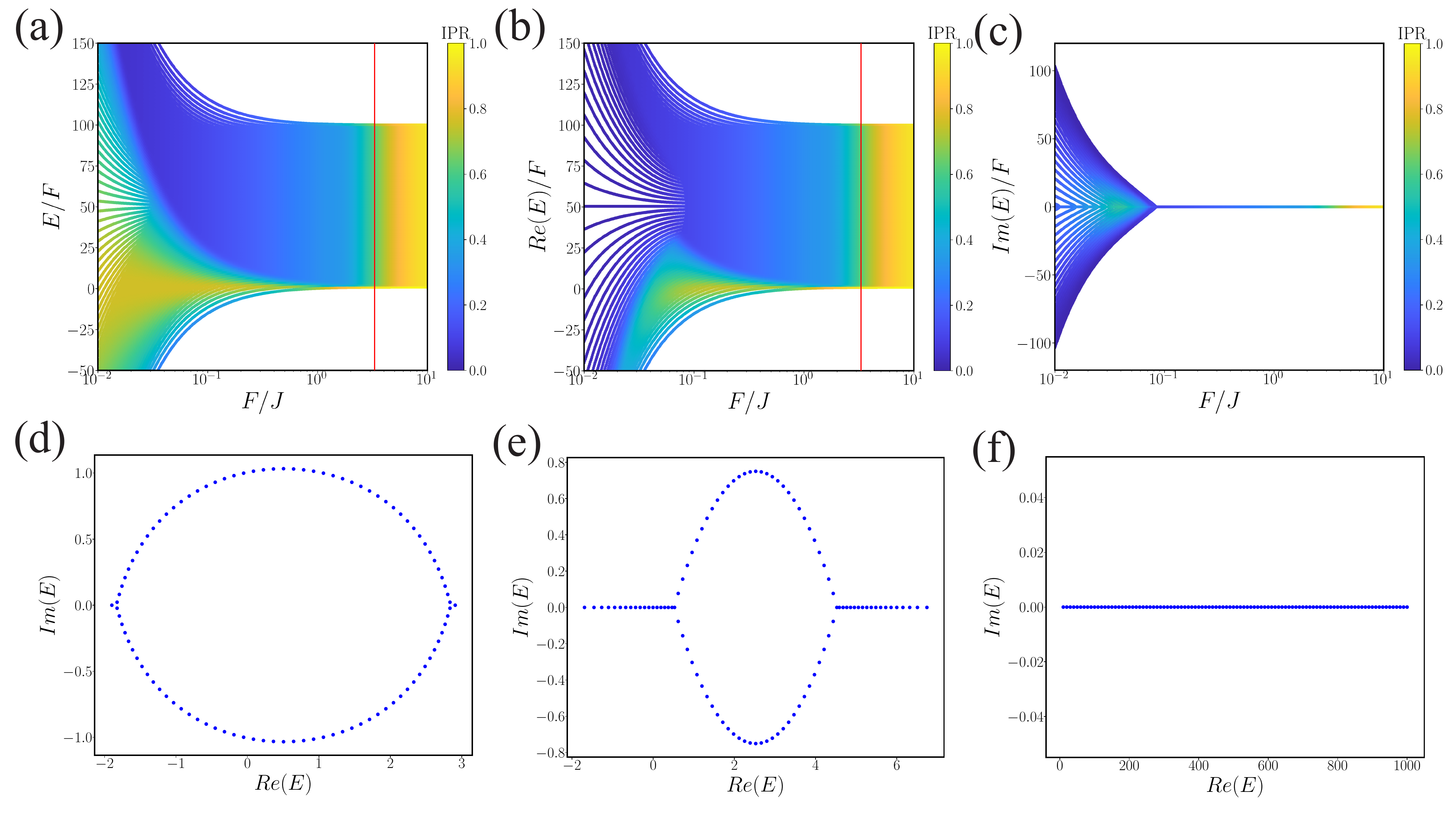}
\caption{The $\rm{IPR}$ and the energy spectrum versus $F/J$ for non-Hermitian Stark model with $\kappa = 1$ under the different boundary conditions. The energy spectrum of the system under OBC is shown in (a). While under PBC in (b) and (c), the system has real-complex transitions, where (b) and (c) represent the real and imaginary parts of the energy spectrum. The red lines in (a) and (b) denote the localization transition point $F_{c2}/J \simeq 3.3$. (d)-(f) The real and imaginary energy spectrum in the complex plane for non-Hermitian stark model under PBC with $F = 0.01$ (d), $0.05$ (e), and $10$ (f). Here, the system size $L = 100$ and $g=0.5$.} 
\label{IPRNH_ka_1}
\end{figure*}

The relation between the eigenstates of $H$ and $H^{\prime}$ is achieved naturally: $|\Psi\rangle=S^{-1}|\Psi^{\prime} \rangle$. Here $H|\Psi\rangle=E|\Psi\rangle$ and the $|\Psi\rangle=\sum_n \psi_{n}|n\rangle$ is the eigenstate of $H$. The transformation
$S^{-1}$ can convert the extended states $|\Psi^{\prime}\rangle$ into skin states, which exponentially gather the wave function all to one of boundaries. For a given localized state of $H^{\prime}$, it may be expressed as $|\psi_{n}^{\prime}|\propto e^{-\gamma|n-n_{0}|}$. $n_{0}$ is the localization center of a particle and $\gamma$ is the LE of the localized state for the Hamiltonian $H^{\prime}$. Then the corresponding wave function of $H(g\neq0)$ may take the following form:
\begin{equation}\label{localization}
|\psi_n|\propto  
\left\{
\begin{array}{cc}
e^{-(\gamma(E)-g)(n-n_0)},&n>n_0, \\
e^{-(\gamma(E)+g)(n_0-n)}, &n<n_0, \\
\end{array}
\right.
\end{equation}
which exhibits different localization length on the different sides of the localization center. When $|g|\geq\gamma$, delocalization occurs on one side and the transition point
from the localized state to the skin state is given by
\begin{equation}\label{gamma}
	\gamma(E)=|g|.
\end{equation}
Through the above discussions, the transition of the existence-nonexistence of NHSE under OBC and the real-complex transition under PBC are analytically determined by
\begin{equation}\label{skin}
\left|a_{\kappa}F_{c1}\right|=\frac{2e^{\kappa |g|+1}}{L}.
\end{equation}

To determine the ME boundaries that separate the delocalized and localized eigenstates, we could exclude the constant $\ln(\kappa N)-1$ in $ \gamma(E)$ as it can be absorbed by wave function normalization. Since a localized eigenstate is not affected by the boundary condition of the system, it then follows that the boundary of the Stark localization-delocalization transition under PBC is also determined by Eq. (\ref{gamma}). Hence, the non-Hermitian ME can be given by
\begin{equation}\label{starklocalization}
\left|a_{\kappa}F_{c2}\right|=2e^{\kappa |g|}.
\end{equation}
The Eq. (\ref{skin}) describes the existence-nonexistence of NHSE and the Eq. (\ref{starklocalization}) determines the boundary of the localization-delocalization transition, which are the central results of this work.

Now, we consider the non-Hermitian mosaic Stark lattice with the $\kappa=1$ and $g= 0.5$. There is no ME in this case. From Eqs. (\ref{a}) and (\ref{skin}), we know $a_{\kappa = 1}=1$ and the critical electric field of the real-complex transition is $\left|F_{c1}\right|=\frac{2e^{|g|+1}}{L}$, which is size-dependent. This indicates that when the electric field $F<F_{c1}$, the NHSE exists, while $F>F_{c1}$, the NHSE will disappear, suggesting that the NHSE is size-dependent, which is different from the general NHSE. Besides, we can also obtain the Stark localization transition point $\left|F_{c2}\right|=2e^{|g|}$ from Eq. (\ref{starklocalization}), which shows that non-Hermiticity from nonreciprocal hopping can suppress the Stark localization and enhance the critical strength of the linear onsite potential.
\begin{figure}[htb]
\includegraphics[width=0.48\textwidth]{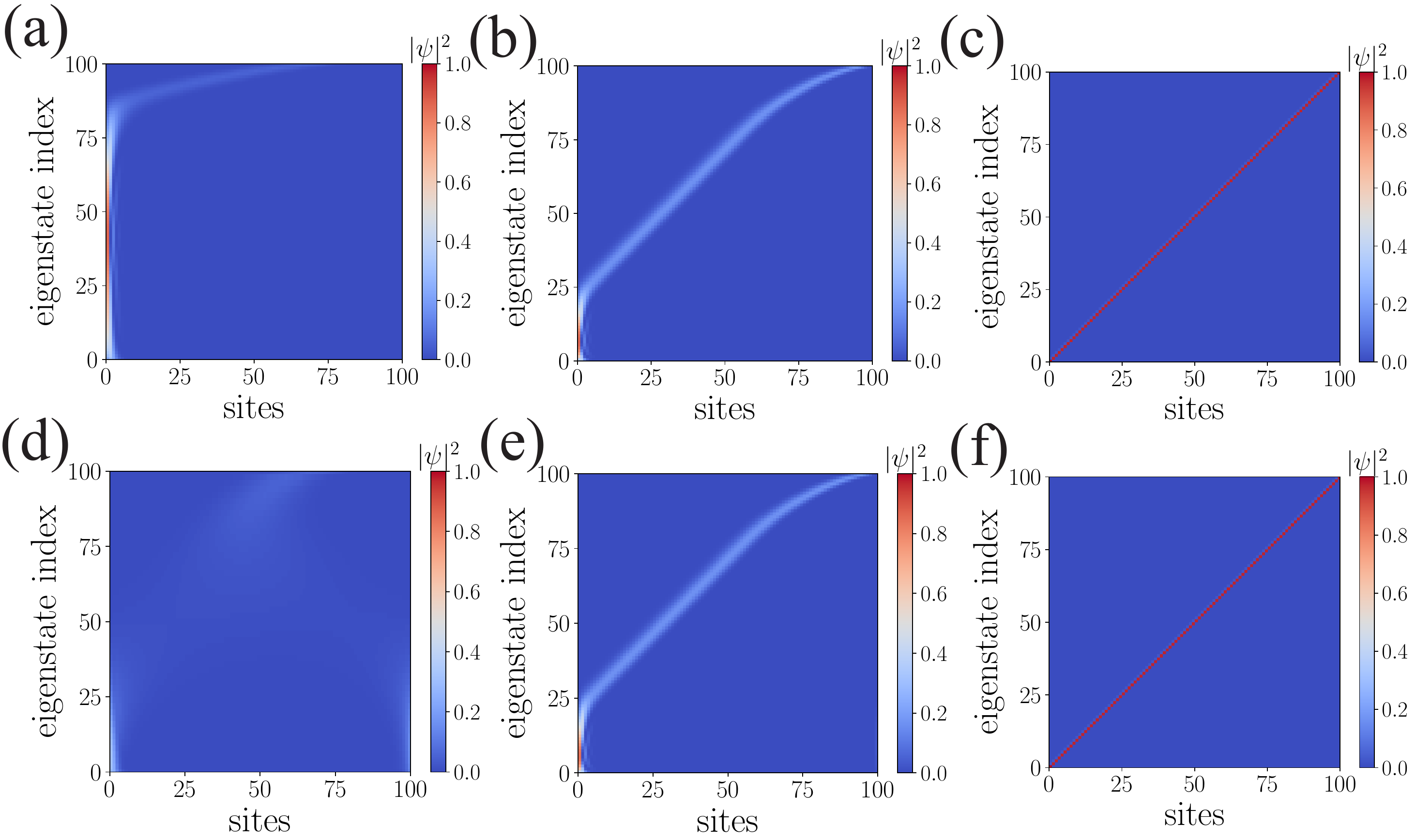}
\caption{The density distribution of all eigenstates for non-Hermitian Stark model with $\kappa = 1$. (a)-(c) represent OBC case, while (d)-(f) represent PBC case. The two rows from left to right corresponds to $F/J = 0.01$, $0.1$, and $10$. Here, the system size $ L = 100$ and $g = 0.5$.}
\label{DiseigNH_ka_1}
\end{figure}

As shown in Fig. \ref{IPRNH_ka_1}(a), we plot $\rm IPR$ and the energy spectrum versus $F/J$ under OBC where the system only has real energy spectrum. When $F/J$ is relatively small, there are not only extended states, but also skin states which are localized at the boundary, where we plot the density distribution of the eigenstates under OBC when $F/J=0.01$. The system can undergo a phase transition from mixed phase to localized phase at transition point $F_{c2}/J = 2e^{0.5}\simeq 3.3$, which is denoted by the red solid line. In Figs. \ref{IPRNH_ka_1}(b) and (c), we show the real and imaginary parts of the energy spectrum under PBC versus $F/J$, respectively. For this model, the boundary effects are weakened and skin states vanish due to PBC when $F/J$ is small. In Figs. \ref{IPRNH_ka_1}(d)-(f), we exhibit the energy spectrum of non-Hermitian Stark system in the complex plane under PBC from left to right corresponding to the case of $F/J = 0.01$, $0.05$ and $10$. We find that when the system is in localized phase, the energy spectrum is real. 

In Fig. \ref{DiseigNH_ka_1}, we show the density distribution of all eigenstates under OBC (top raw) and PBC (bottom raw) from left to right corresponding $F/J = 0.01$, $0.1$ and $10$. In Figs. \ref{IPRNH_ka_1}(a) and \ref{DiseigNH_ka_1}(a), we can find the particle is localized at the boundary, which is caused by the NHSE under OBC when $F/J = 0.01$. As the $F/J$ increases, the skin states reduce and the system becomes fully localized when $F/J>3.3$ as shown in Fig. \ref{IPRNH_ka_1}(c). Moreover, we can see all eigenstates are extended obeying the Bessel functions which are different with the disorder or quasiperiodic systems when $F/J = 0.01$ in Fig. \ref{DiseigNH_ka_1}(d). Similarly, the particle becomes fully localized when $F/J = 10$ as shown in Fig. \ref{DiseigNH_ka_1}(f). Thus, the non-Hermitian Stark systems without mosaic structure still have fully localized phases. 

\begin{figure}[tb]
\includegraphics[width=0.4\textwidth]{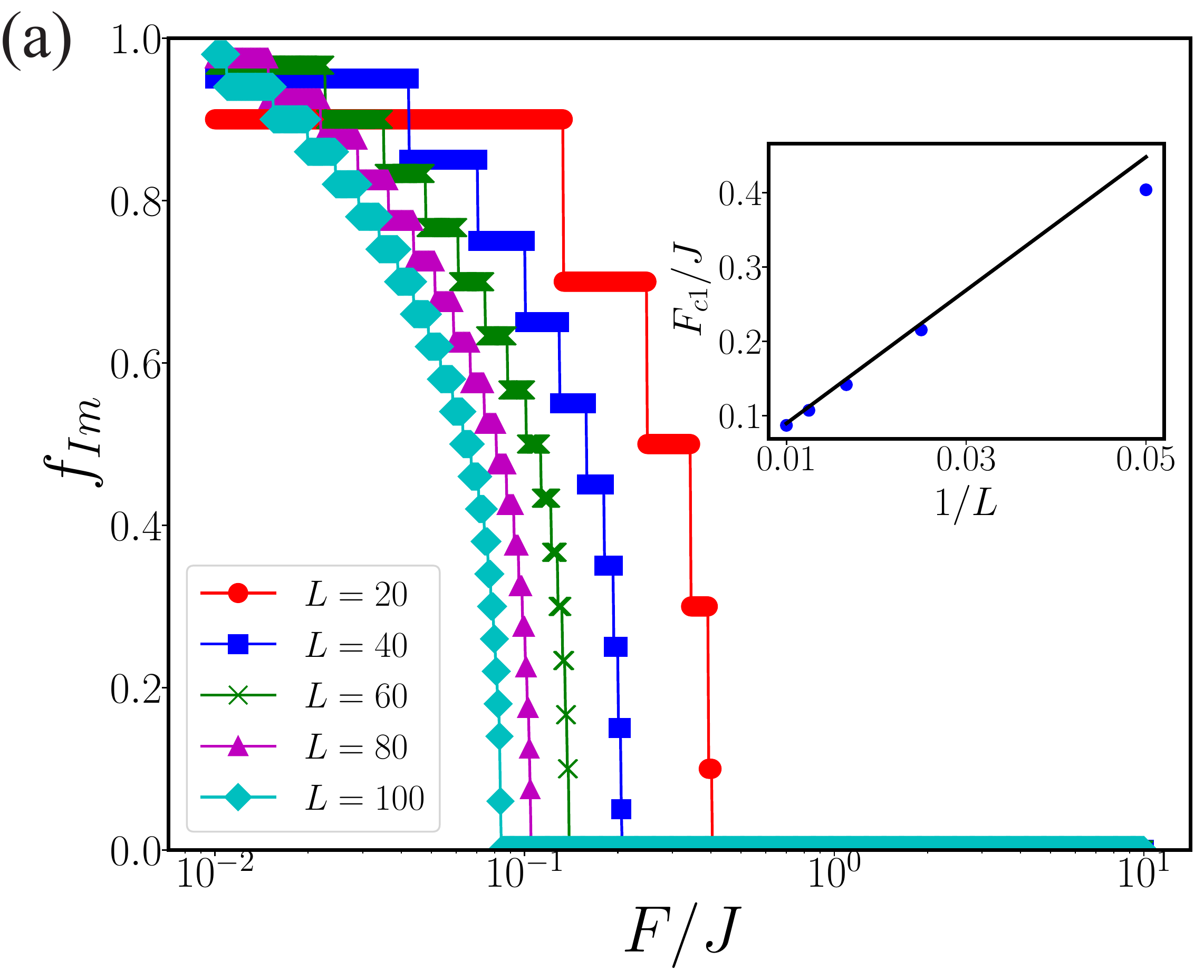}
\includegraphics[width=0.425\textwidth]{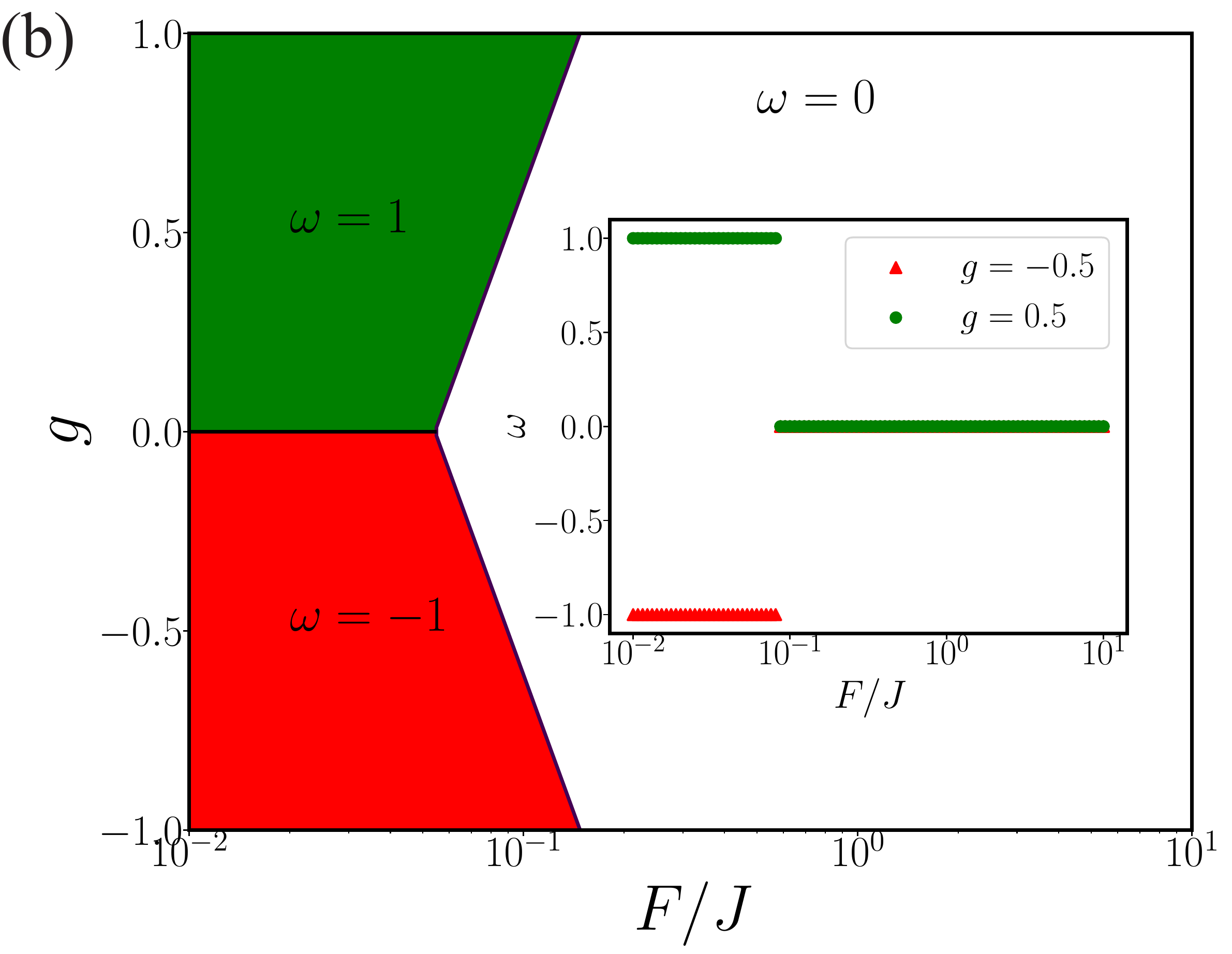}
\caption{(a) $f_{Im}$ versus $F/J$ for $L=20, 40, 60, 80$, and $100$ for the non-Hermitian Stark model with $\kappa =1$ under PBC. The real-complex transition point for the different system size is shown in the inset. The blue dots are numerical results, and the blank solid line is the analytical results of Eq. (\ref{skin}). (b) The spectral winding number $\omega$ of non-Hermitian Stark model with $\kappa = 1$ and the system size $L =100$. $\omega = \pm 1 (\omega = 0)$ correspond to the existence (nonexistence) of NHSE, respectively. The inset is the numerical results of the $\omega$ versus $F/J$ for $g = 0.5$ (green) and $g = -0.5$ (red). }
\label{real_winding}
\end{figure}

From Eq. (\ref{skin}), we know the transition of the existence-nonexistence of NHSE is always accompanied by the real-complex transition under PBC. The critical electric field is inversely proportional to the lattice length, giving $F_{c1}\propto 1/L$. We define a quantity $f_{Im} = D_{Im}/D$ to character the proportion of the complex energy spectrum, where $D_{Im}$ represents the number of the eigenenergies with nonzero imaginary part and $D$ is the system size. Here, if the imaginary part of a eigenenergy is less than $10^{-13}$, we neglect the imaginary part of this energy and see it as a real energy. In Fig. \ref{real_winding}(a), we show the numerical results of the real-complex transition depend on different lattice size $L$. We find the critical point of the real-complex transition decreases with the increasing of $L$. The numerical results of the real-complex transition points $F_{c1}$ versus $1/L$ are depicted in the inset of Fig. \ref{real_winding}(a). These results of the real-complex transitions are consistent with the analytical result Eq. (\ref{skin}).

\begin{figure*}[t]
\includegraphics[width=0.8\textwidth]{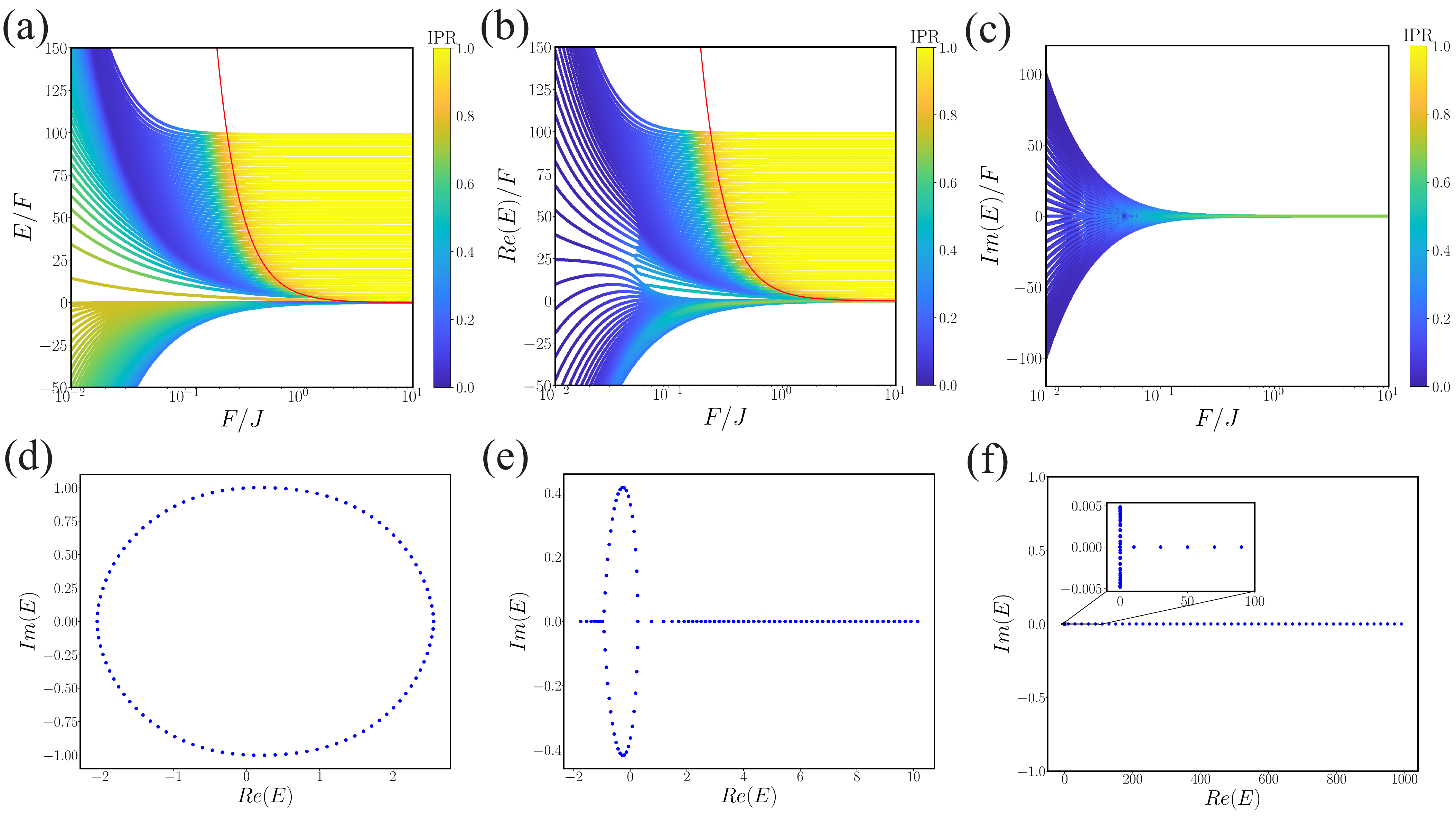}
\caption{The $\rm{IPR}$ and energy spectrum versus $F/J$ for non-Hermitian mosaic Stark model with $\kappa = 2$ under different boundary conditions. The energy spectrum of the system under OBC is shown in (a). While under PBC, the energy spectrum of system is complex. The $\rm{IPR}$ corresponding to real and imaginary part of eigenenergies spectrum are drawn in (b) and (c). The red solid lines represent ME. (d)-(f) The real and imaginary energy spectrum in the complex plane for non-Hermitian mosaic stark model under PBC with $F = 0.01$ (d), $0.1$(e), and $10$(f). The half of the eigenenergies are complex and the other half are real when $F=10$ as shown in the inset of (f). Here, the system size $L = 100$ and $g=0.5$.}
\label{NHIPRka_2}
\end{figure*}

Moreover, in order to calculate the winding number $\omega$ versus $F/J$ and $g$ for non-Hermitian Stark model with $\kappa = 1$, we introduce the non-Hermitian ring with a magnetic flux $\Phi$ under the twist boundary condition, i.e., $H(\Phi) = H + J e^{g+i\Phi}c_L^{\dagger}c_1 +J e^{-g-i\Phi}c_1^{\dagger}c_L$ \cite{jiang2019interplay, peng2022manipulating,liu2021localization}. The winding number $\omega$ is defined as
\begin{eqnarray}\label{winding1}
\omega =\frac{1}{2\pi i}\int_0^{2\pi} \text{d}\Phi\partial_{\Phi}\ln \det[ H(\Phi)-E_{B} ],
\end{eqnarray}
where $E_B$ is the prescribed basis point which is not an eigenenergy of $H(\Phi)$. In this way, we can diagnose the transition of the existence-nonexistence of NHSE under OBC. As shown in Fig. \ref{real_winding}(b), $\omega=\pm 1$ ($\omega=0$) correspond to the existence  (nonexistence) of NHSE, respectively. The inset of Fig. \ref{real_winding}(b) gives the numerical results based on Eq. (\ref{winding1}) which is consistent with the analytical result Eq. (\ref{skin}).

\section{Non-Hermitian mobility edge without disorder}\label{section4}
In this section, we investigate the non-Hermitian ME in the Hamiltonian described by Eq. (\ref{model}) with $\kappa=2$ and $g=0.5$. Based on the Eq. (\ref{starklocalization}) and  $a_{\kappa = 2}=E$ from the Eq. (\ref{a}), we obtain the energy-dependent non-Hermitian ME as the following form  
\begin{equation}\label{starklocalization_2}
\left|EF_c\right|=2e.
\end{equation}
The eigenenergies of extended states for the model (\ref{model}) lie in ${\rm{Re}}(E)<E_{c}$, whereas eigenenergies of the localized states lie in ${\rm{Re}}(E)>E_{c}$. When the Stark localization is emerging, the absolute value of the maximum energy $\left|E_{max}\right|=LF$, then we take it into Eq.(\ref{starklocalization_2}) and obtain $F_c=\sqrt{2e/L}$.

\begin{figure}[b]
\includegraphics[width=0.48\textwidth]{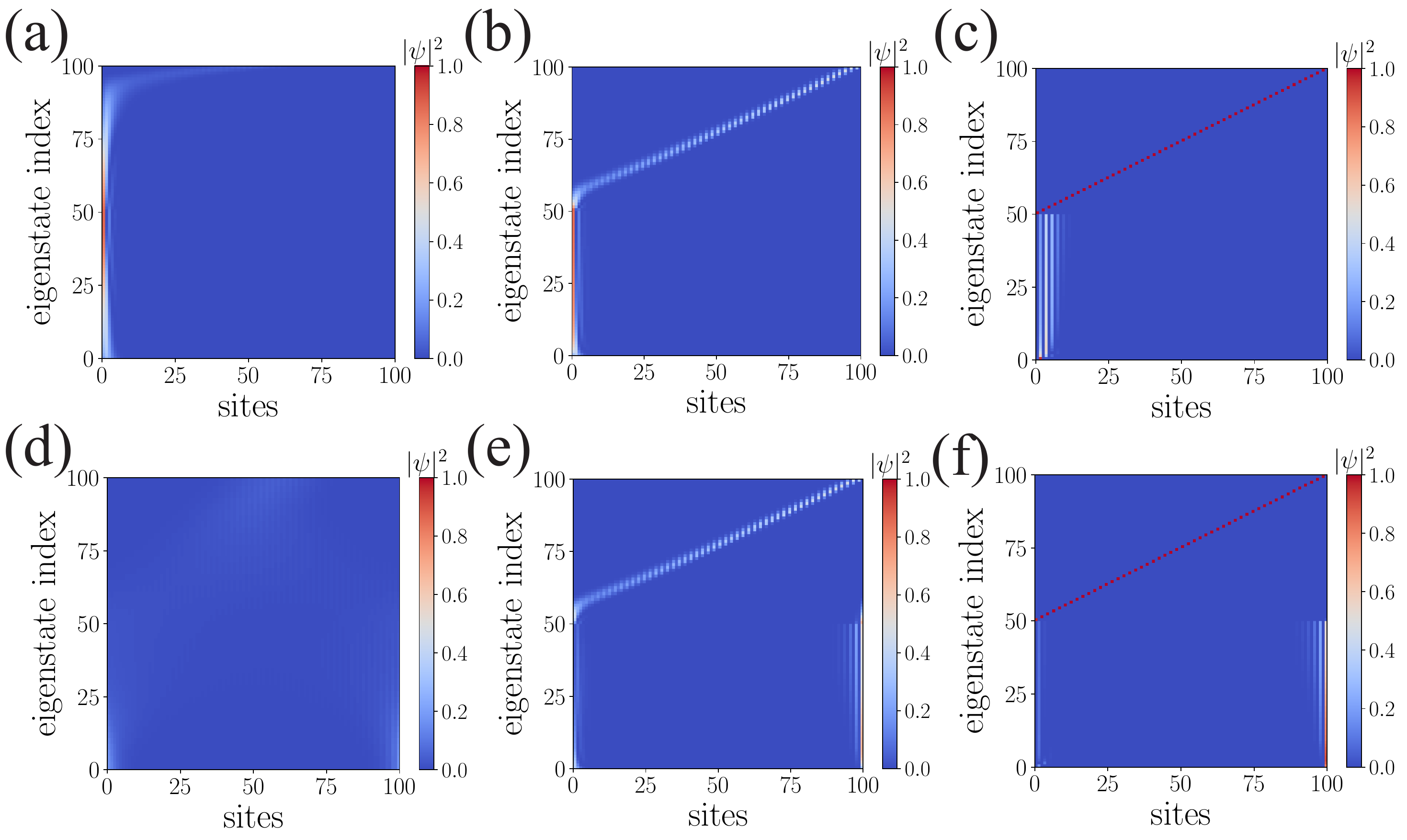}
\caption{The density distribution of all eigenstates for non-Hermitian Stark model with $\kappa = 2$. (a)-(c) represent OBC case, while (d)-(f) represent PBC case. The two rows from left to right corresponds to $F/J = 0.01$, $0.1$, and $10$. Here, the system size $ L = 100$ and $g = 0.5$.}
\label{DiseigNH_ka_2}
\end{figure}

We consider the non-Hermitian mosaic Stark model with $\kappa =2$ and the system size $L= 100$. We first plot the $\rm{IPR}$ and the energy spectrum versus $F/J$ under OBC in Fig. \ref{NHIPRka_2}(a). In this case, this system only has real energy spectrum. While for the case of PBC, all eigenenergies have imaginary parts under very weak fields as shown in Fig. \ref{NHIPRka_2}(d). As the electric field increases and exceeds a critical value $F_c =\sqrt{2e/100}\simeq 0.23$, the system enters a mixed phase with a ME, where the number of localized and extended states will be divided by half. We plot the real and imaginary parts of the energy spectrum under PBC versus $F/J$ in Figs. \ref{NHIPRka_2}(b) and (c), respectively. We also show the energy spectrum in the complex plane under PBC for the case of $F/J =0.01$, $0.1$ and $10$ in Figs. \ref{NHIPRka_2}(d)-(e). We find that when $F/J=10$, the imaginary part of the eigenenergies is very small in the ME phase. We show the density distribution of all eigenstates for the non-Hermitian Stark model with $\kappa = 2$ under OBC in Figs. \ref{DiseigNH_ka_2} (a)-(c) and PBC in Figs. \ref{DiseigNH_ka_2} (d)-(e) from left to right corresponding to $F/J = 0.01, 0.1$ and 10. We find that the system exists the NHSE under OBC in Fig. \ref{DiseigNH_ka_2}(a) when $F/J = 0.01$, while under PBC, the NHSE disappears in Fig. \ref{DiseigNH_ka_2}(d). As the $F/J$ increases, the skin effects do not vanish under OBC in Fig. \ref{DiseigNH_ka_2}(c), in contrast to the gigantic field can restrain the NHSE in the $\kappa=1$ case. As shown in Figs. \ref{NHIPRka_2}(f) and \ref{DiseigNH_ka_2}(f), the half of the eigenstates are fully localized and the others are extended under PBC when $F/J = 10$. Thus, the non-Hermitian mosaic Stark lattice (\ref{model}) with $\kappa = 2$ only exists the mixed phase with ME for the strong field.

\begin{figure}[b]
\includegraphics[width=0.48\textwidth]{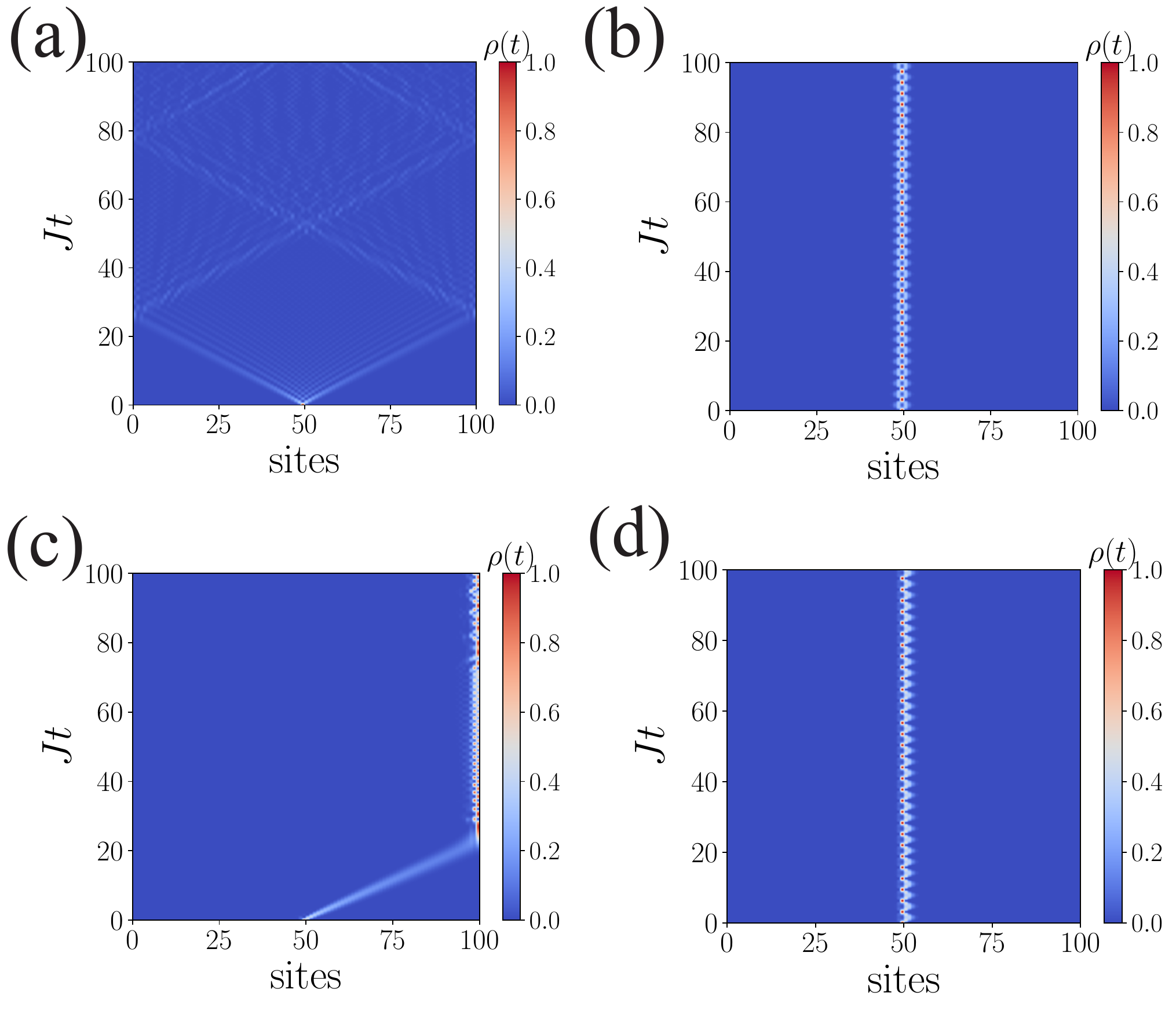}
\caption{Time evolution of the particle density at any site for Stark model with $\kappa = 1$ under OBC. (a) and (b) represent the Hermitian case of  $F/J = 0.01$ and $2$ for $g = 0$, while (b) and (c) represent the non-Hermitian case of $F/J = 0.01$ and $2$ for $g = 0.5$. Here, the system size $L=100$.}
\label{rhot_ka_1}
\end{figure}
\section{Expansion Dynamics}\label{section5}
\begin{figure}[b]
\includegraphics[width=0.48\textwidth]{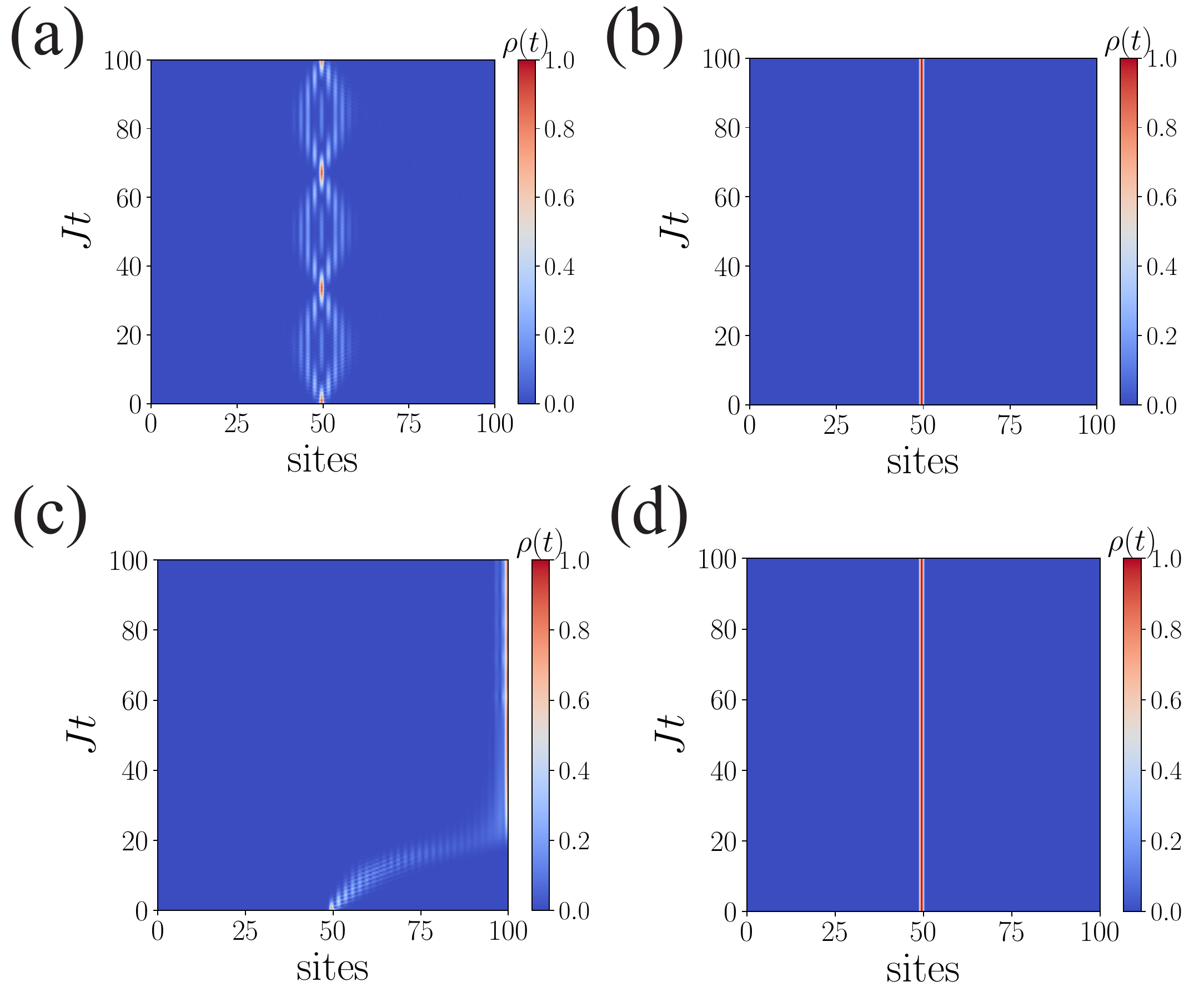}
\caption{Time evolution of the particle density at any site for mosaic Stark model with $\kappa = 2$ under OBC. (a) and (b) represent the Hermitian case of $F/J = 0.1$ and $10$ for $g =0$, while (c) and (d) represent the non-Hermitian case of $F/J=0.1$ and $10$ for $g = 0.5$. Here, the system size $L=100$. }
\label{rhot_ka_2}
\end{figure}

To dynamically probe the properties of NHSE and the existence of MEs in the model (\ref{model}), we also study the expansion dynamics of the wave function \cite{li2022dynamic,zhai2022nonequilibrium,orito2022unusual,xu2020dynamical,xu2021dynamical}. We first consider the initial state $|\Psi(0) \rangle = | n_0\rangle$ in which only one particle is initially at the center site $n_0$ of the lattice. The time evolution states are determined by 
\begin{equation}
|\Psi(t)\rangle = \frac{e^{-iHt}|\Psi(0)\rangle}{\left\vert \left\vert e^{-iHt}|\Psi(0)\rangle \right\vert \right\vert}.
\end{equation}
So we can get the particle density at any site and time $t$,
\begin{equation}
\rho_n(t) = |C_n(t)|^2,   
\end{equation} 
where $C_n = \langle n| \Psi(t)\rangle$ is the expansion coefficient of the time evolution states, and $|n\rangle$ is $n$-th computational basis of the Hilbert space. In the following, we consider the case of different system parameters under OBC. 
We calculate the time evolution of the probability density for the Stark model with $\kappa = 1$ and $\kappa = 2$ in Figs. \ref{rhot_ka_1} and \ref{rhot_ka_2} under OBC, respectively. We show the particle density distribution of the case of $g = 0$ and $F/J = 0.01$ in Fig. \ref{rhot_ka_1}(a) and $F/J = 2$ in Fig. \ref{rhot_ka_1}(b). This system presents a Bloch oscillation in the extended region. When $F/J = 2$, the single particle oscillates around the center of the lattice. With the increasing of $F/J$, the particle will locate at the center of the lattice. As the result of the non-Hermitian Stark model with $\kappa = 1$ under OBC in Fig. \ref{rhot_ka_1}(c), we find that the particle will move to the right boundary in the late time evolution when $F/J = 0.01$. This phenomenon is induced by the NHSE. When $F/J = 2$, we also find the Bloch oscillation in Fig. \ref{rhot_ka_1} (d). However, compared to Fig. \ref{rhot_ka_1} (b), the particle asymmetrically oscillates around the center and biases toward the right boundary due to the non-Hermiticity. Similarly, we still find that Bloch oscillation can happen in the Hermitian mosaic Stark model ($\kappa = 2$) when $F/J = 0.1$ as shown in Fig. \ref{rhot_ka_2} (a), whereas it is not observed for the non-Hermitian mosaic Stark model with $\kappa = 2$. We dynamically observe the NHSE of the non-Hermitian mosaic Stark system with $\kappa = 2$ when $F/J = 0.1$ in Fig. \ref{rhot_ka_2} (c). Then, when $F/J = 10$, both the Hermitian and the non-Hermitian systems are in the mixed phase with ME. The particle is localized at the initial site in Figs. \ref{rhot_ka_2}(b) and (d).

\begin{figure}[tb]
\includegraphics[width=0.4\textwidth]{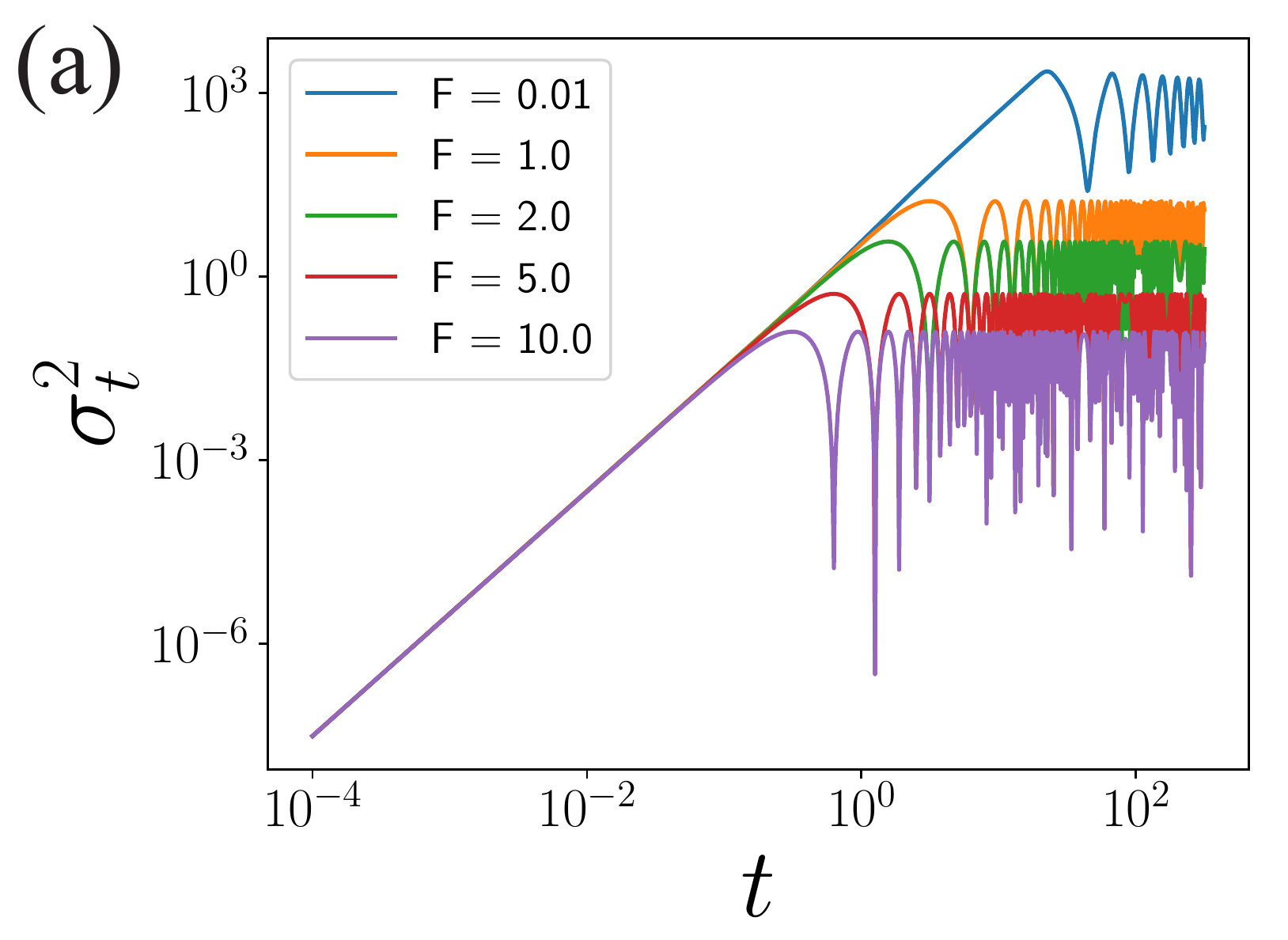}
\includegraphics[width=0.4\textwidth]{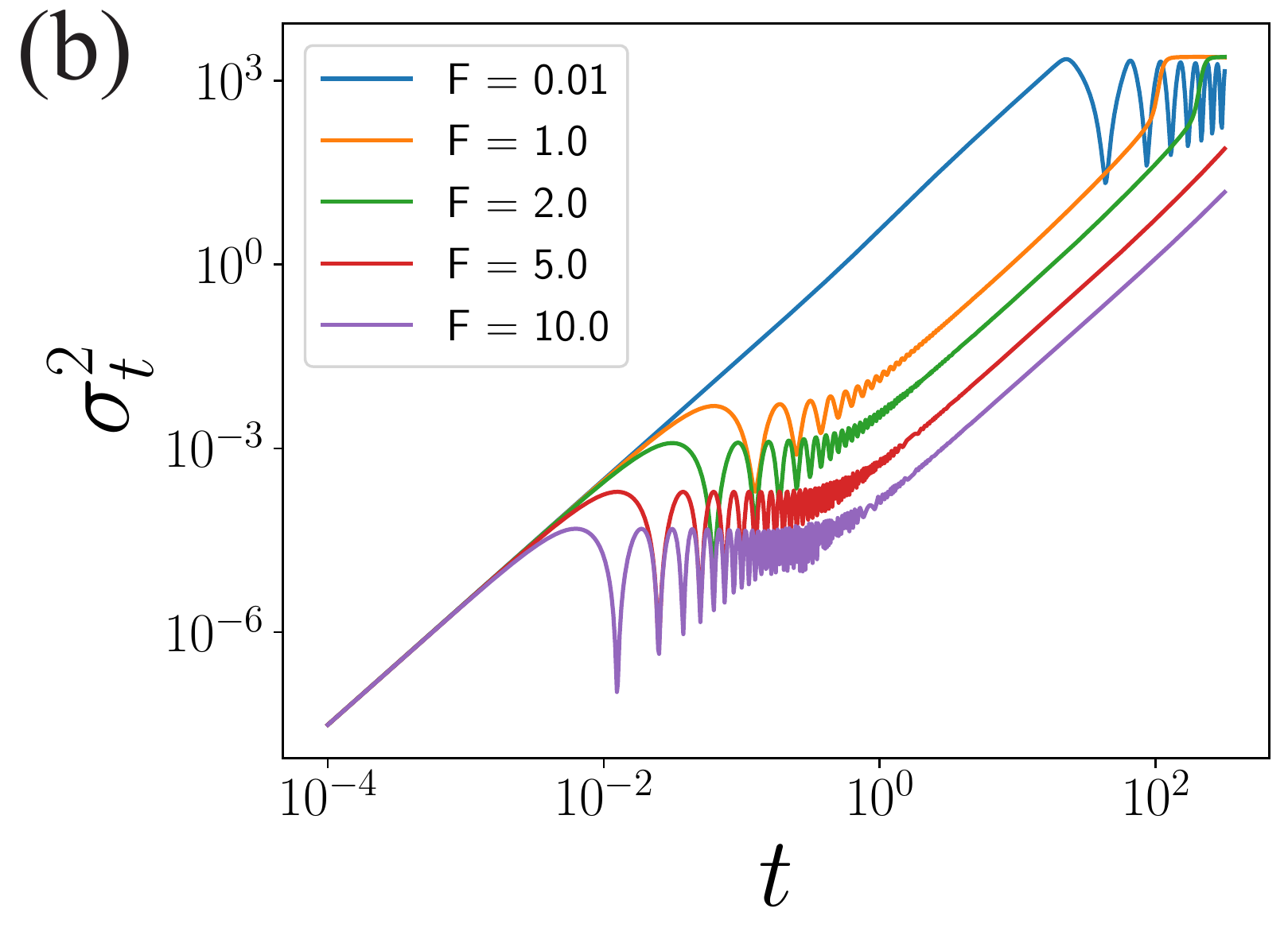}
\caption{The time evolution of the mean-square displacement for non-Hermitian ($g= 0.5$) Stark model with $\kappa =1$ (a) and $2$ (b) under PBC, respectively. Here, the system size $L = 100$ and $J=1$.}
\label{sigmat}
\end{figure}

We calculate the mean-square displacement $\sigma^2_t$ to estimate the spreading of the width of a wave packet \cite{xu2020dynamical,xu2021dynamical,li2023multiple,qi2023multiple}, which is defined as 
\begin{equation}
    \sigma^2_t = \sum_n |n - n_0|^2 |C_n(t)|^2.
\end{equation}
Here, we show the case of the non-Hermitian Hamiltonian $H(g = 0.5)$ with $\kappa = 1$ and $2$ under PBC in Figs. \ref{sigmat} (a) and (b), respectively. We find $\sigma^2_t$ increases directly to $10^3$ until $t \approx 10$ when $F/J = 0.01$ corresponding to the extended phase for the above two case. For the non-Hermitian Stark model with $\kappa =1$, the threshold of $\sigma_t^2$ decrease and the time to threshold reduce with the increasing of $F/J$. For the non-Hermitian Stark model with $\kappa = 2$, the $\sigma^2_t$ have a saturation value when the system is in the ME phase ($F/J = 1.0, 2.0, 5.0$, and $10.0$) approximately between $t = 10^{-2}$ and $t = 1$ which is dependent on $F/J$. The $\sigma_t^2$ will increase to around $10^3$ due to the existence of the single particle mobility edge with the time evolution when $t> 1$. These numerical results indicate the dynamical signatures of the existence of the
NHSE and ME in the non-Hermitian disorder-free lattices.

\section{Conclusion}\label{sec:conclusion}
We have investigated the fate of the NHSE and the ME in a non-Hermitian mosaic Stark lattice with electric fields. For the $\kappa = 1$ case, we find that a critical electric field which describes the existence-nonexistence of the size-dependent NHSE is given by $F_{c1} \propto 1/L$. The result indicates that a weak field is sufficient to suppress the NHSE in the thermodynamic limit. When the electric field strength is strong and exceeds the critical value $F_{c2}$, the Wannier-Stark ladder forms and the localization induced by the field win the competition with NHSE. Thus, NHSE vanishes and the system is in the Stark localized phase.
For the $\kappa = 2$ case and the weak field, the non-Hermitian ME emerges and the localized states earlier appear. In the strong field regime, the skin states are still existent and stable compared to the $\kappa = 1$ case. Moreover, we also confirm the existence of the NHSE and the non-Hermitian ME from the dynamical evolution of an initial wave packet and unveil that they display the different behaviors in the different regimes. Our work could expand the understanding of the NHSE and the non-Hermitian ME in 1D disorder-free systems. For future studies, on the theory side, the non-Hermitian Stark localization and ME can be extended to the  many-body disorder-free system~\cite{zhang2021mobility,wei2022static,li2023non,liu2023ergodicity}. On the experimental side, our results can be directly simulated and verified in  recent electrical circuit experiments~\cite{jiang2019interplay,peng2022manipulating,helbig2020generalized,liu2021non,zhang2023electrical,su2023simulation}.

\section{Acknowledgments}
The financial supports from National Key R$\&$D Program of China (Grant No. 2021YFA1402104), National Natural Science Foundation of China (Grant No. 12074410, No. 12047502, No. 11934015 and No. 12247103) and Strategic Priority Research Program of the Chinese Academy of Sciences (Grant No. XDB33000000) are gratefully acknowledged.

\bibliography{Localization}
\end{document}